\documentclass[11pt]{article}
\usepackage{Blois,epsfig}
\usepackage{amssymb}
\usepackage{amsfonts}

\def\be{\begin{equation}}
\def\ee{\end{equation}}
\def\bea{\begin{eqnarray}}
\def\eea{\end{eqnarray}}

\newcommand{\alphaEM}{\alpha} 

\newcommand{\pert}{P} 
\newcommand{\nprt}{N\!P} 
\newcommand{\pbar}{{\bar{p}}}

\newcommand{\benn}{\begin{displaymath}} 
\newcommand{\eenn}{\end{displaymath}} 
\newcommand{\beann}{\begin{eqnarray*}} 
\newcommand{\eeann}{\end{eqnarray*}} 
\newcommand{\barray}{\begin{array}} 
\newcommand{\earray}{\end{array}} 
\newcommand{\inv}{\frac{1}}

\newcommand{\fm}{\mbox{fm}} 
\newcommand{\mb}{\mbox{mb}}

\newcommand{\GeV}{\mbox{GeV}} 
\newcommand{\TeV}{\mbox{TeV}} 
\newcommand{\G}{{\cal G}}       
\newcommand{\GG}{\hat{\cal{G}}} 
\newcommand{\Identity}{{1\!\rm l}}

\newcommand{\Pc}{{\cal P}}      
\newcommand{\Ps}{{\cal P}_S}    
\newcommand{\Tr}{\mbox{Tr}}    

\begin{document}
\vspace*{2cm}
\begin{center}
\Large{\textbf{XIth International Conference on\\ Elastic and Diffractive Scattering\\ Ch\^{a}teau de Blois, France, May 15 - 20, 2005}}
\end{center}

\vspace*{2cm}
\title{THE HEIDELBERG POMERON}

\author{ H.J. PIRNER}

\address{University of Heidelberg, Institute of Theoretical Physics,\\
Philosophenweg 19, 69120 Heidelberg, Germany}

\maketitle


One of the challenges in quantum chromodynamics (QCD) is the  
description and understanding of hadronic high-energy scattering.  
Since the momentum transfers can be small, the QCD coupling constant  
is too large for a reliable perturbative treatment. Non-perturbative  
QCD is needed which is rigorously only available as a computer  
simulation on Euclidean lattices. 
  
An interesting phenomenon observed in hadronic high-energy scattering  
is the rise of the total cross sections with increasing c.m.\ energy.  
While the rise is slow in hadronic reactions of {\em large} particles  
such as protons, pions, kaons, or real photons, it  
is steep if only one {\em small} particle is involved such as an  
incoming virtual photon or an  
outgoing charmonium.  
    
In this work, we develop a model combining perturbative and  
non-perturbative QCD to compute high-energy reactions of hadrons and  
photons with special emphasis on saturation effects that manifest the  
$S$-matrix unitarity. Aiming at a unified description of  
hadron-hadron, photon-hadron, and photon-photon reactions involving  
real and virtual photons as well, we follow the {\em functional  
  integral approach} to high-energy scattering in the eikonal  
approximation,  
in which the $S$-matrix element factorizes into the universal  
correlation of two light-like Wegner-Wilson loops $S_{DD}$ and  
reaction-specific light-cone wave functions. The light-like  
Wegner-Wilson loops describe color-dipoles given by the quark and  
antiquark in the meson or photon and in a simplified picture by a  
quark and diquark in the baryon. Consequently, hadrons and photons are  
described as color-dipoles with size and orientation determined by  
appropriate light-cone wave  
functions. Thus, the {\em  
  loop-loop correlation function} $S_{DD}$ is the basis for our  
unified description.  
 We compute the $T$-matrix in a {\em functional  
  integral approach} developed for parton-parton  
scattering in the {\em eikonal approximation}.  In this  
approach, the $T$-matrix element for the reaction $ab \rightarrow cd$  
factorizes as follows  
\bea    
     &&   T_{ab \rightarrow cd}(s,t) =  
        2is \int \!\!d^2b_{\!\perp}   
        e^{i {\vec q}_{\!\perp} {\vec b}_{\!\perp}}  
        \int \!\!dz_1 d^2r_1 \!\int \!\!dz_2 d^2r_2             
        \nonumber \\      
     &&\times   \psi_c^*(z_1,{\vec r}_1)\,\psi_d^*(z_2,{\vec r}_2)  
        \left[1-S_{DD}({\vec b}_{\!\perp},z_1,{\vec r}_1,z_2,{\vec r}_2)\right]  \psi_a(z_1,{\vec r}_1)\,\psi_b(z_2,{\vec r}_2)   
        \ , \label{Eq_model_T_amplitude} \eea   
where the {\em loop-loop correlation function}   
\be  
        S_{DD}({\vec b}_{\!\perp},z_1,{\vec r}_1,z_2,{\vec r}_2)  
        = \Big\langle W[C_1] W[C_2] \Big\rangle_G  
\label{Eq_loop_loop_correlation_function}  
\ee  
describes the elastic scattering of two color-dipoles (DD) with  
transverse size and orientation ${\vec r}_i$ and longitudinal quark  
momentum fraction $z_i$ at impact parameter ${\vec b}_{\!\perp}$,  
transverse momentum transfer ${\vec q}_{\!\perp}$ ($t = -{\vec  
  q}_{\!\perp}^{\,\,2}$) and c.m.\ energy squared $s$. 
  
\begin{figure}[p]
\setlength{\unitlength}{1.cm}  
  \begin{center}  
        \epsfig{file=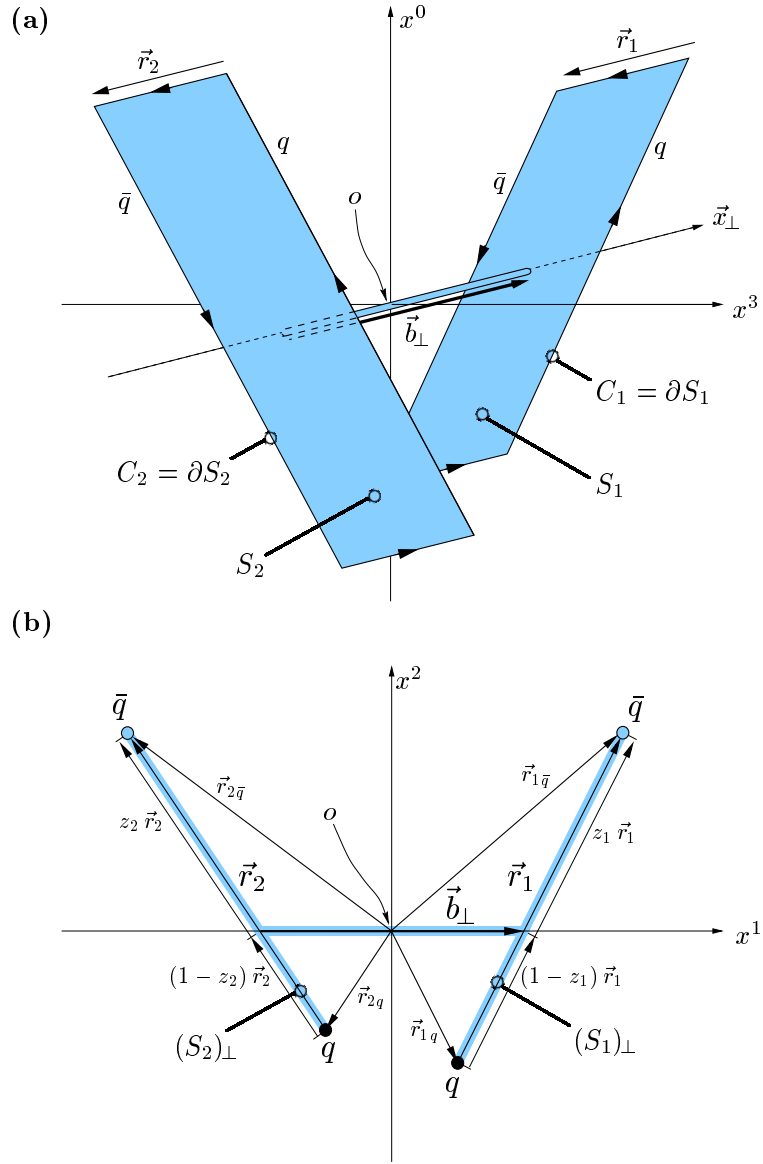,width=10.cm}  
  \end{center}  
\caption{\small Space-time (a) and transverse (b) arrangement of the Wegner-Wilson loops.}  
\label{Fig_loop_loop_scattering_surfaces}  
\end{figure}

\begin{figure}[p]  
\setlength{\unitlength}{1.cm}  
\begin{center}  
\epsfig{file=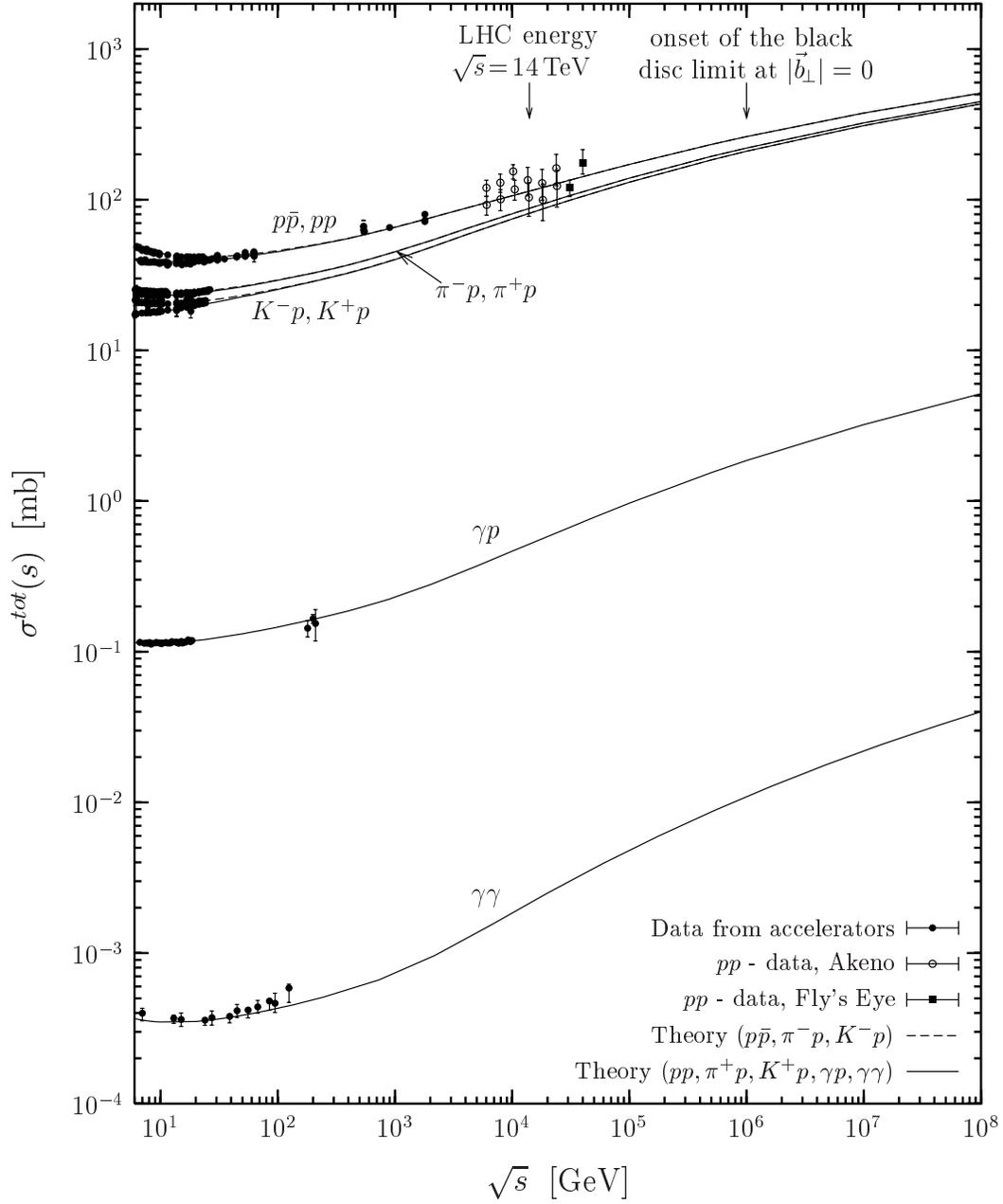,width=14cm}  
\end{center}  
\caption{\small The total cross section $\sigma^{tot}$ is shown as a  
  function of the c.m.\ energy $\sqrt{s}$ for $pp$, $p\pbar$,  
  $\pi^{\pm}p$, $K^{\pm}p$, $\gamma p$ and $\gamma \gamma$ scattering.  
  The solid lines represent the model results for $pp$, $\pi^+p$, $K^+p$,  
  $\gamma p$ and $\gamma \gamma$ scattering and the dashed lines the  
  ones for $p\pbar$, $\pi^-p$, and $K^-p$ scattering. The $pp$,  
  $p\pbar$, $\pi^{\pm}p$, $K^{\pm}p$, $\gamma p$  
  and $\gamma \gamma$ data taken at  
  accelerators are indicated by the closed circles while the closed  
  squares (Fly's eye data) and the open  
  circles (Akeno data) indicate cosmic ray data.  
  The arrows at the top point to the LHC energy, $\sqrt{s} =  
  14\,\TeV$, and to the onset of the black disc limit in $pp$ ($p\pbar$)  
  reactions, $\sqrt{s} \approx 10^6\,\GeV$.}  
\label{Fig_sigma_tot}  
\end{figure}

The path of each color-dipole is represented by a {\em light-like QCD  
  Wegner-Wilson loop}  
\be  
        W[C_{1,2}] =   
        \inv{N_c} \Tr\,\Pc  
        \exp\!\left[-i g\!\oint_{\scriptsize C_{1,2}}\!\!dz^{\mu}  
        \G_{\mu}(z) \right]        
        \ ,  
\label{Eq_Wegner-Wilson_loop}  
\ee  
where $N_c$ is the number of colors, $\Tr$ the trace in color space,  
$g$ the strong coupling, and $\G_{\mu}(z) = \G_{\mu}^a(z) t^a$ the  
gluon field with the $SU(N_c)$ group generators $t^a$ that demand the  
path ordering indicated by $\Pc$. Quark-antiquark dipoles are represented by loops in the fundamental  
$SU(N_c = 3)$ representation. In the eikonal approximation to  
high-energy scattering the $q$ and ${\bar q}$ paths form straight  
light-like trajectories.  
Figure~\ref{Fig_loop_loop_scattering_surfaces} illustrates the  
space-time (a) and transversal (b) arrangement of these loops. 
The QCD vacuum expectation value $\langle \ldots \rangle_G$ in the  
loop-loop correlation function  represents functional  
integrals.
To compute the loop-loop correlation function, we transform the line  
integrals over the loops $C_{1,2}$ into integrals over surfaces  
$S_{1,2}$ with $\partial S_{1,2} = C_{1,2}$ by applying the {\em  
  non-Abelian Stokes'  
theorem} and  
the cumulant expansion truncated at $n=2$  
 Consequently, all higher cumulants, $K_n$ with  
$n>2$, vanish\footnote{We are going to use the cumulant expansion in  
  the Gaussian approximation also for perturbative gluon exchange.  
  Here certainly the higher cumulants are non-zero.} and the loop-loop  
correlation function can be expressed in terms of $K_2$  
\bea  
&& \!\!\!\!\!\!\!\!\!\!\!  
        \Big\langle W[C_{1}] W[C_{2}] \Big\rangle_G   
        \nonumber \\  
&& \!\!\!\!\!\!\!\!\!\!\!  
        = \inv{N_c^2} \Tr_2  
        \exp\!\left[-\frac{g^2}{8} \!  
          \int_{S} \! d\sigma^{\mu\nu}(x_1) \!  
          \int_{S} \! d\sigma^{\rho\sigma}(x_2)   
          \Big\langle \Ps\,   
          \GG_{\mu\nu}(o,x_1;C_{x_1 o})   
          \GG_{\rho\sigma}(o,x_2;C_{x_2 o})  
          \Big\rangle_G   
        \right]  
        . \nonumber \\     
\label{Eq_matrix_cumulant_expansion_<W[C1]W[C2]>}  
\eea  
Due to the color-neutrality of the vacuum, the gauge-invariant bilocal  
gluon field strength correlator simplifies 
\be  
        \Big\langle  
        \frac{g^2}{4\pi^2}\,  
        \G^a_{\mu\nu}(o,x_1;C_{x_1 o})  
        \G^b_{\rho\sigma}(o,x_2;C_{x_2 o})  
        \Big\rangle_G  
        =: \inv{4}\delta^{ab}   
        F_{\mu\nu\rho\sigma}(x_1,x_2,o;C_{x_1 o},C_{x_2 o}) \ .  
\label{Eq_Ansatz}  
\ee  
 The  
quantity $F_{\mu\nu\rho\sigma}$ is obtained from the stochastic vacuum model. We define  
\be  
        \chi_{S_i S_j}  
        := - \, i \frac{\pi^2}{4}   
        \int_{S_i} \! d\sigma^{\mu\nu}(x_1)   
        \int_{S_j} \! d\sigma^{\rho\sigma}(x_2)  
        F_{\mu\nu\rho\sigma}(x_1,x_2,o;C_{x_1 o},C_{x_2 o}) \ ,  
\label{Eq_chi_SS}          
\ee  
then    
\bea  
        \Big\langle W[C_{1}] W[C_{2}] \Big\rangle_G &=&   
        \inv{N_c^2} \Tr_2  
        \exp\!\Bigg[-i\, \frac{1}{2}   
                \Big\{\,  
                \left(\chi_{S_1 S_2}+\chi_{S_2 S_1}\right)  
                \big(t^a \,\otimes\, t^a\big)   
        \nonumber \\  
        &&  + \,\chi_{S_1 S_1}   
                \big(t^a t^a \,\otimes\, \Identity\big)   
            + \chi_{S_2 S_2}  
                \big(\Identity \,\otimes\, t^a t^a\big)   
                \,\Big\}\,  
              \Bigg]  
        \ .   
\label{Eq_eikonal_functions_<W[C1]W[C2]>}  
\eea  
Our ansatz for the tensor structure of $F_{\mu\nu\rho\sigma}$  
leads to $\chi_{S_1 S_1} = \chi_{S_2  
  S_2} = 0$ for light-like loops, and also to 
$\chi_{S_1 S_2} =  \chi_{S_2 S_1} =: \chi$.  For the evaluation of the trace of the  
remaining exponential, we project on the $\bar 3$ and $6$  
representations and obtain:  
\be         
        \Big\langle W[C_{1}] W[C_{2}] \Big\rangle_G  
         =  \frac{2}{3}  
        \exp\!\left[-  \frac{i}{3}\chi \right]  
        +  
        \frac{1}{3}  
        \exp\!\left[\frac{2i}{3}\chi \right]   
        \   
\label{Eq_final_result_<W[C1]W[C2]>}  
\ee  
We decompose the gauge-invariant bilocal gluon field strength  
correlator~(\ref{Eq_Ansatz}) into a perturbative ($\pert$) and  
non-perturbative ($\nprt$) component  
\be  
        F_{\mu\nu\rho\sigma}   
        = F_{\mu\nu\rho\sigma}^{\nprt} + F_{\mu\nu\rho\sigma}^{\pert}   
        \ .  
\label{Eq_F_decomposition}  
\ee  
Here, $F_{\mu\nu\rho\sigma}^{\nprt}$ gives the low frequency  
background field contribution modelled by the non-perturbative {\em  
  stochastic vacuum model} (SVM) and  
$F_{\mu\nu\rho\sigma}^{\pert}$ the additional high frequency  
contributions described by {\em perturbative gluon exchange}.  Such a  
decomposition is supported by lattice QCD computations of the  
Euclidean field strength  
correlator.  
Note that $\chi = \chi_{c}^{\nprt} + \chi_{nc}^{\nprt} + \chi^{\pert}$ is  
a real-valued function. Since, in addition, the wave functions  
$|\psi_i(z_i,\vec{r}_i)|^2$ used in this work are invariant under the replacement  
$(\vec{r}_i \rightarrow -\vec{r}_i, z_i \rightarrow 1-z_i)$, the  
$T$-matrix element becomes purely imaginary and reads for $N_c=3$ (for details we refer to the literature [3,4,5,6,7]) 
\bea  
        \!\!\!\!\!\!\!\!\!\!\!\!\!  
        T(s,t)   
        & = & 2is \int \!\!d^2b_{\!\perp}   
                e^{i {\vec q}_{\!\perp} {\vec b}_{\!\perp}}  
                \int \!\!dz_1 d^2r_1 \!  
                \int \!\!dz_2 d^2r_2 \,\,  
                |\psi_1(z_1,\vec{r}_1)|^2   \,\,  
                |\psi_2(z_2,\vec{r}_2)|^2         
        \nonumber \\      
        && \!\!\!\!\!\!\!\!\!\!  
        \times   
        \left[1-\frac{2}{3}   
        \cos\!\left(\frac{1}{3}  
        \chi({\vec b}_{\!\perp},z_1,\vec{r}_1,z_2,\vec{r}_2)\!\right)  
        - \frac{1}{3}  
        \cos\!\left(\frac{2}{3}  
        \chi({\vec b}_{\!\perp},z_1,\vec{r}_1,z_2,\vec{r}_2)\!\right)  
        \right].  
\label{Eq_model_purely_imaginary_T_amplitude}  
\eea  
The good agreement of the computed total cross sections with the  
experimental data is shown in Fig.~\ref{Fig_sigma_tot}.  
Here, the solid lines represent the theoretical results for $pp$,  
$\pi^+p$, $K^+p$, $\gamma p$, and $\gamma \gamma$ scattering and the  
dashed lines the ones for $p\pbar$, $\pi^-p$, and $K^-p$ scattering.  
The $pp$, $p\pbar$, $\pi^{\pm}p$, $K^{\pm}p$, $\gamma  
p$ and $\gamma \gamma$  
data taken at accelerators are indicated by  
the closed circles while the closed squares (Fly's eye  
data) and the open circles (Akeno  
data) indicate cosmic ray data. Concerning the  
photon-induced reactions, only real photons are considered.  
  
The prediction for the total $pp$ cross section at LHC ($\sqrt{s} =  
14\,\TeV$) is $\sigma^{tot}_{pp} = 114.2\,\mb$ in good agreement with  
the cosmic ray data. Compared with other works, our LHC prediction is  
close to the one of Block et al.,  
$\sigma^{tot}_{pp} = 108 \pm 3.4\,\mb$, but considerably larger than  
the one of Donnachie and Landshoff,  
$\sigma^{tot}_{pp} = 101.5\,\mb$.  
  
The differences between $ab$ and $\bar{a}b$ reactions for $\sqrt{s}  
\lesssim 100\,\GeV$ result solely from the different reggeon  
contributions which die out rapidly as the energy increases. The  
pomeron contribution to $ab$ and $\bar{a}b$ reactions is, in  
contrast, identical and increases as the energy increases. It thus  
governs the total cross sections for $\sqrt{s} \gtrsim 100\,\GeV$ where  
the results for $ab$ and $\bar{a}b$ reactions coincide.  
  
The differences between $pp$ ($p\pbar$), $\pi^{\pm}p$, and $K^{\pm}p$  
scattering result from the different transverse extension parameters,  
$S_p = 0.86\,\fm > S_{\pi} = 0.607\,\fm > S_{K} = 0.55\,\fm$.  Since a smaller transverse  
extension parameter favors smaller dipoles, the total cross section  
becomes smaller, and the short distance physics described by the  
perturbative component becomes more important and leads to a stronger  
energy growth due to $\epsilon^{\pert} = 0.73 > \epsilon^{\nprt} =  
0.125$. In fact, the ratios $\sigma^{tot}_{pp}/\sigma^{tot}_{\pi p}$  
and $\sigma^{tot}_{pp}/\sigma^{tot}_{Kp}$ converge slowly towards  
unity with increasing energy as can already be seen in  
Fig.~\ref{Fig_sigma_tot}.  
For real photons, the transverse size is governed by the constituent  
quark masses $m_f(Q^2=0)$,  
where the light quarks have the strongest effect, i.e.\   
$\sigma^{tot}_{\gamma p} \propto 1/m_{u,d}^2$ and  
$\sigma^{tot}_{\gamma \gamma} \propto 1/m_{u,d}^4$. Furthermore, in  
comparison with hadron-hadron scattering, there is the additional  
suppression factor of $\alphaEM$ for $\gamma p$ and $\alphaEM^2$ for  
$\gamma \gamma$ scattering coming from the photon-dipole transition.  
In the $\gamma \gamma$ reaction, also the box diagram  
contributes but is neglected  
since its contribution to the total cross section is less than  
1\%.

To conclude we give a list of references to the work done in our group 
on high-energy scattering during the last years: Ref.~[1] derives the 
Hamiltonian for a $q \bar q $-state on the light cone from the 
calculation of a single Wegner-Wilson loop near the light cone. A 
consistent
derivation of the wave functions and the scattering is very important.
Ref.~[2] presents arguments for geometrical scaling  and 
calculations that the energy 
dependence of high-energy scattering is related to critical correlations 
[6,8]
of Wilson lines approaching the light cone in an effective  
$2+1$-dimensional QCD
which is simulated in Ref.~[9] on the lattice.
Refs.~[3,10] calculate from the stochastic vacuum model the constant and 
$\log(1/x)$ term in the gluon distribution, which are used in Ref.~[3] 
as input to a DGLAP-calculation of the proton structure function. 
Refs.~[4,5,7,12]  contain calculations closely related to the subject of 
this talk. Ref.~[4] relates the loop-loop correlation in Euclidean
space to the loop-loop correlation in Minkowski space. Ref.~[5] describes 
how the nonperturbative string can be decomposed into many perturbative 
dipoles which allow to calculate the unintegrated gluon distribution.
Ref.~[7] 
explains the contents of this talk in great detail,
it also contains further references to other work  left out in this 
talk. Ref.~[12] discusses the specific behavior of the dipole-proton 
cross 
section for large dipoles in the stochastic 
vacuum model and calculates vector-meson resonance production cross 
sections. Resonances have larger sizes than groundstate hadrons 
and are therefore sensitive to the cross section for large dipoles. 
Ref.~[11] makes the case for a soft and hard Pomeron picture of 
the proton structure function using
the stochastic model for the soft pomeron and perturbation theory for 
the hard pomeron.

This work owes a great deal to my colleagues H.G. Dosch and O. Nachtmann 
in Heidelberg who have developed the basics of the loop-loop correlation 
model and with whom I have had the pleasure to collaborate.

\end{document}